\documentclass[a4paper,12pt]{article}

\usepackage{latexsym,amsmath,amsfonts,amssymb}
\usepackage{multirow}
\usepackage[dvips]{graphicx}
\usepackage{mathrsfs,mathtools}
\usepackage{pstricks,pst-node,pst-text,pst-3d}
\usepackage{slashed}
\usepackage{verbatim}
\usepackage{feynmp}
\usepackage[nosort]{cite}
\usepackage[bulletsep]{collref}

\newcommand{\tr}[1]{\text{Tr}\big[#1\big]}

\newcommand{\sect}[1]{\setcounter{equation}{0}\section{#1}}

\begin{document}

\begin{titlepage}

\setcounter{page}{0}

\begin{flushright}
{ }
\end{flushright}

\vspace{0.6cm}

\begin{center}
  {\Large \bf SL(2) sector: weak/strong coupling agreement of three-point correlators}

\vskip 0.8cm

{\bf George Georgiou}\footnote{georgiou@inp.demokritos.gr}
\\
{\sl
Demokritos National Research Center \\ Institute of Nuclear Physics\\
Ag. Paraskevi, GR-15310 Athens, Greece}\\

\vskip 1.2cm

\end{center}

\begin{abstract}
We evaluate three-point correlation functions of single trace operators in $N=4$ SYM
at both weak and strong coupling. We focus on the case where two of the operators
belong in a $SL(2)$ sub-sector, and are dual to string solutions in a broad
class of solutions with large $S$ and $J$ charges, while the third operator is a BPS state.
Perfect agreement between the  structure constants at weak and strong coupling is found.
Finally, comments on this matching, as well as on the space-time structure of the correlators, are given.
\end{abstract}

\vfill

\end{titlepage}

\sect{Introduction}\label{sec:intro}

${\cal N}=4$ Super Yang-Mills  (SYM) theory is an
important example of an interacting four dimensional Conformal Field Theory (CFT)
which has been
thoroughly studied because of the AdS/CFT duality with string
theory~\cite{Maldacena:1998re,Witten:1998qj}. Furthermore, it is the
first  interacting four dimensional theory that we have significant chances to
completely "solve". Being a conformal field theory "solving" it means
to be, at least, able to identify its primary operators and their conformal dimensions.
The second necessary piece of information needed is the structure constants
which determine the Operator Product Expansion (OPE) between two
primary operators.
Recently, huge progress has been made in the computation of the planar contribution
to the conformal dimensions of non-protected
operators for any value of the coupling constant,
using integrability (for a recent review see \cite{Beisert:2010jr}).
On the other hand, very little is known about the
structure constants.
The aim of this note is to contribute towards this second direction.



Our current knowledge of the OPE coefficients is essentially based on
a perturbative expansion around $\lambda=0$, where standard Feynman
diagrams can be used to evaluate the relevant gauge theory
correlators, or around $\lambda=\infty$ where the IIB string theory is
well approximated by a simpler description.  By comparing the 3-point
correlators among half-BPS operators in these two different limits,
the authors of~\cite{Lee:1998bxa} conjectured that the corresponding
structure constants are non-renormalised (i.e. they have a trivial
dependence on the 't~Hooft coupling).  On the contrary the 3-point
correlators among non-protected operators receive quantum corrections,
as it is shown, for instance, by the correlator between three Konishi
operators~\cite{Bianchi:2001cm}. On the gauge theory side, the authors
of~\cite{Beisert:2002bb,Chu:2002pd,Okuyama:2004bd,Roiban:2004va,Alday:2005nd,Alday:2005kq,Grossardt:2010xq,Georgiou:2009tp}
studied systematically the structure constants and computed the corrections arising from the planar
1-loop Feynman diagrams. The importance of the operator mixing for the operators
participating in the correlators was stressed in \cite{Georgiou:2008vk,Georgiou:2009tp,Georgiou:2011xj}.
On the string theory side it is more
difficult to extract information about non-protected OPE coefficients,
since, in the supergravity limit, all non-protected operators acquire
large conformal dimension and decouple. The BMN
limit~\cite{Berenstein:2002jq} represents a different approximation,
where it is possible to extract useful information on non-BPS
structure constants.

Recently, another approach to the calculation of n-points correlators
involving non-BPS states was developed
\cite{Yoneya:2006td,Dobashi:2004nm,Tsuji:2006zn,Janik:2010gc,Buchbinder:2010vw,Zarembo:2010rr,Costa:2010rz}.
More precisely, the authors of \cite{Janik:2010gc} argued that it should be possible to
obtain the correlation functions of local operators corresponding to classical
spinning string states, at strong coupling, by evaluating the string action on
a classical solution with appropriate boundary conditions
after convoluting  with the relevant to the classical states wavefunctions.
In \cite{Buchbinder:2010vw,Roiban:2010fe,Ryang:2010bn,Klose:2011rm}, 2-point and 3-point correlators of vertex operators representing
classical string states with large  spins were calculated.
Finally, in a series of papers
\cite{Zarembo:2010rr,Costa:2010rz,Hernandez:2010tg,Russo:2010bt,Georgiou:2010an,Park:2010vs,Bak:2011yy,Bissi:2011dc,Hernandez:2011up,Ahn:2011zg,Arnaudov:2011wq,Ahn:2011dq,Arnaudov:2011ek} the 3-point function coefficients
of a correlator involving a massive string state, its conjugate and a third "light" state
state were computed. This was done by taking advantage of the known classical solutions
corresponding to the 2-point correlators of operators dual to massive string states.

More recently, three-point functions of single trace operators were studied in \cite{Escobedo:2010xs,Escobedo:2011xw}
from the perspective of integrability. In particular, an intriguing weak/strong coupling match of correlators
involving operators in the $SU(3)$ sector was observed \cite{Escobedo:2011xw}. This match was found to hold
for correlators of two non-protected operators in the Frolov-Tseytlin limit and one short BPS operator.
In this note, we extend this weak/strong coupling match for correlators involving operartors in the $SL(2,R)$
closed subsector of $N=4$ SYM theory. One important difference with respect to the $SU(3)$ case is that our string solutions
are not point-like in the $AdS_5$ space which means that their field theory duals have a large number of covariant derivatives
along a light-like direction.

The plan for the rest of this paper is as follows.
In Section 2, we give a short review of classical string solutions in the Frolov-Tseytlin limit having one large spin $S$ in
$AdS_5$ and one large spin $J$ in $S^5$. Subsequently, we write down the analytically continued version of these solutions
describing the propagation of a string
extending along a light-like direction on the boundary of the $AdS_5$
into the bulk and back to the boundary. We then proceed and use
the formalism of \cite{Zarembo:2010rr} to evaluate
the three-point function coefficient, at strong coupling, of a correlator involving
two of the aforementioned $SL(2,R)$ operators and a BPS state.
In Section 3, we evaluate the same three-point function coefficient at weak coupling
by employing a coherent state description of the SL(2,R) operators valid in the limit we consider.
Perfect agreement between the weak and strong coupling result is found.
It should be noted that the agreement we find is valid for {\it any } solution in the Frolov-Tseytlin limit.
Finally, comments on this matching, as well as on the space-time structure of the aforementioned correlators (see Appendix), are given.

\sect{Strong coupling regime}
\label{sec:2-pointgen}
In this Section, we give a short review of classical string solutions in the Frolov-Tseytlin limit having one large spin $S$ in
$AdS_5$ and one large spin $J$ in $S^5$. These solutions are the string counterparts
of single trace gauge invariant operators belonging in the $SL(2)$ closed subsector \footnote{The anomalous dimension of twist 2 operators
has been studied extensively, both at weak coupling \cite{Callan:1973,Gross:1974cs,Axenides:2002zf,Kotikov:2004er}
for theories with different amounts of supersymmetry
and at
strong coupling \cite{Gubser:2002tv,Frolov:2002av,Basso:2007wd,Kruczenski:2007cy}
for the maximally supersymmetric theory.} of the full $PSU(2,2|4)$
algebra of $N=4$ supersymmetric Yang-Mills (SYM) theory. The operators we will be focusing on can be written schematically as
\begin{equation}\label{twistJop}
{\cal O}_{SJ}=\tr{D^S_+Z^J}+...
\end{equation}
where $Z$ is one of the complex scalars of $N=4$ SYM and $D_+$ is the covariant derivative along a light-cone
direction.\\
Here we follow closely \cite{Bellucci:2004qr,Stefanski:2004cw}.
As mentioned above, the string states we are interested in have two non-zero
charges, one with respect to one of the isometries of $AdS_5$ and the other with respect to one of the isometries of the
five-sphere $S^5$. Thus, it is enough to consider solutions embedded in a $AdS_3 \times S^1$ subspace of the full
$AdS_5 \times S^5$ manifold. The metric of this subspace reads
\begin{equation}\label{metric}
ds^2=-\cosh^2{\tilde\rho} \, dt^2+d\tilde\rho^2+\sinh^2{\tilde\rho} \, d{\tilde\phi}_1^2+d{\gamma}_1^2.
\end{equation}
The bosonic part of the Polyakov action can be written as
\begin{eqnarray}\label{Polyakov}
S=\frac{\sqrt{\lambda}}{4 \pi}\int d\sigma d\tau g_{\mu\nu}(\partial_{\tau}X^{\mu}\partial_{\tau}X^{\nu}-
\partial_{\sigma}X^{\mu}\partial_{\sigma}X^{\nu}).
\end{eqnarray}
Besides satisfying the equations of motion that follow from \eqref{Polyakov} the solutions must satisfy the
Virasoro constraints
\begin{eqnarray}\label{Virasoro}
g_{\mu\nu}\partial_{\tau}X^{\mu}\partial_{\sigma}X^{\nu}=0 \nonumber\\
g_{\mu\nu}(\partial_{\tau}X^{\mu}\partial_{\tau}X^{\nu}+
\partial_{\sigma}X^{\mu}\partial_{\sigma}X^{\nu}).
\end{eqnarray}
Subsequently, one can  employ the change of variables
\begin{eqnarray}\label{changevar}
{\tilde\phi}_1=\phi+t \qquad \gamma_1=\phi_3-t \qquad {\tilde\rho}=\frac{\rho}{2}
\end{eqnarray}
and look for solutions satisfying the following ansatz
\begin{eqnarray}\label{ansatz}
t=k\tau \qquad \phi=\phi(\sigma,\tau) \qquad \phi_3=\phi_3(\sigma,\tau) \qquad \rho=\rho(\sigma,\tau).
\end{eqnarray}
The Frolov-Tseytlin limit consists in taking
\begin{eqnarray}\label{Fro-tse}
k \rightarrow \infty \,\,\,\,\,\, {\rm while \,\, keeping} \,\,\,\,\,\,k \partial_{\tau}X^{\mu}={\rm finite}, \,
\partial_{\sigma}X^{\mu}={\rm finite} \qquad X^{\mu}=\rho,\phi,\phi_3.
\end{eqnarray}
Then to leading order in $k$ the first Virasoro constraint \eqref{Virasoro} becomes
\begin{eqnarray}\label{Virasoro1}
k\big( (\cosh{\rho}-1) \partial_{\sigma}\phi-2\partial_{\sigma}\phi_3 \big)=0.
\end{eqnarray}
This equation can be used to eliminate $\partial_{\sigma}\phi_3$ in terms of $\partial_{\sigma}\phi$.
Under this substitution and at leading in $k$ order, the action \eqref{Polyakov} takes the form
\begin{eqnarray}\label{action2}
S=\frac{\sqrt{\lambda}}{4 \pi}\int d\sigma d\tau \Big( k \big( (\cosh{\rho}-1) \partial_{\tau}\phi-2\partial_{\tau}\phi_3 \big)
-\frac{1}{4} \big( (\partial_{\sigma}\rho)^2 +\sinh^2{\rho} (\partial_{\sigma}\phi)^2\big) \Big).
\end{eqnarray}
Changing the variable from $\tau$ to t and introducing the effective coupling $\lambda'=\frac{\lambda}{J^2}$ ($J\approx \lambda k$ )
we get
\begin{eqnarray}\label{action3}
S=\frac{J}{2}\int_0^{2 \pi} \frac{d\sigma}{2 \pi} \int dt \Big(  \big( (\cosh{\rho}-1) \partial_{t}\phi-2\partial_{t}\phi_3 \big)
-\frac{\lambda'}{4 } \big( (\partial_{\sigma}\rho)^2 +\sinh^2{\rho} (\partial_{\sigma}\phi)^2\big) \Big).
\end{eqnarray}
As shown in \cite{Stefanski:2004cw,Bellucci:2004qr} the same expression appears also on the field theory side
as the effective low energy Lagrangian of the $SL(2)$ spin chain associated to the 1-loop dilatation operator of $N=4$ SYM
restricted to this particular sector. This fact guarantees the agreement, in leading order in $\lambda'$, between the string energies
and the anomalous dimensions of the corresponding operators in field theory \footnote{The agreement between string and field theory
will not continue to hold at arbitrary order in the $\lambda'$ expansion due to the order of limits problem. On the string theory side both
$\lambda$ and $J$ tend to infinity in such a way that $\lambda/J^2$ is small and fixed. On the other hand, on field theory side $\lambda$ (and
thus $\lambda/J^2$) is kept small such that perturbation theory can be applied. Obviously these two limits are not the same.}.

$\bullet$ $\mathbf {Analytic \,\, continuation}$

Although the solution \eqref{changevar}, \eqref{ansatz} is perfectly fine for calculating
the conserved charges of the string, it is not appropriate for calculating
the 2-point function of this solution holographically.
This is because the string of \eqref{changevar}, \eqref{ansatz} lives entirely in the
bulk of $AdS$ and it never touches its boundary.
What we need is a string solution that tunnels from the boundary
of $AdS_5$ to the boundary, in the spirit of \cite{Yoneya:2006td,Dobashi:2004nm}.

This solution can be constructed by performing an analytic continuation to both
the global and world-sheet time. Namely,
\begin{eqnarray}\label{analcont}
t\rightarrow -i t \qquad \tau\rightarrow -i \tau.
\end{eqnarray}
After this the metric \eqref{metric} and the action \eqref{Polyakov}
become
\begin{equation}\label{metricEuc}
ds^2=\cosh^2{\tilde\rho} \, dt^2+d\tilde\rho^2+\sinh^2{\tilde\rho} \, d{\tilde\phi}_1^2+d{\gamma}_1^2.
\end{equation}
\begin{eqnarray}\label{PolyakovEuc}
iS=-S_E=-\frac{\sqrt{\lambda}}{4 \pi}\int d\sigma d\tau g_{\mu\nu}(\partial_{\tau}X^{\mu}\partial_{\tau}X^{\nu}+
\partial_{\sigma}X^{\mu}\partial_{\sigma}X^{\nu}).
\end{eqnarray}
If, as above, we keep only the leading in $k$ terms the expressions for the action and the Virasoro constraints read
\begin{eqnarray}\label{action2Euc}
iS=-S_E=-\frac{\sqrt{\lambda}}{4 \pi}\int d\sigma d\tau \Big( -i k \big( (\cosh{\rho}-1) \partial_{\tau}\phi-2\partial_{\tau}\phi_3 \big)
+\frac{1}{4} \big( (\partial_{\sigma}\rho)^2 +\sinh^2{\rho} (\partial_{\sigma}\phi)^2\big) \Big).\nonumber\\
\end{eqnarray}
\begin{eqnarray}\label{VirasoroEuc}
k\big( (\cosh{\rho}-1) \partial_{\sigma}\phi-2\partial_{\sigma}\phi_3 \big)=0 \nonumber\\
ik\big( (\cosh{\rho}-1) \partial_{\tau}\phi-2\partial_{\tau}\phi_3 \big)=-\frac{1}{4}\big( (\partial_{\sigma}\rho)^2+\sinh^2{\rho}(\partial_{\sigma}\phi)^2 \big)
\end{eqnarray}

In the embedding coordinates, the Euclidean continuation of the solution \eqref{changevar}, \eqref{ansatz} is
\begin{eqnarray}\label{Eucsolution}
Y^{-1}=\cosh{k\tau}\cosh{\tilde\rho},  \qquad Y^0=\sinh{k\tau} \cosh{\tilde\rho},\nonumber \\
Y^1=\cosh{(k\tau+i\phi)}\sinh{\tilde\rho},\,\,\,\,\,\,
Y^4=-i\sinh{(k\tau+i \phi)}\sinh{\tilde\rho},\nonumber \\ \, Y^2=Y^3=0, \qquad
\, {\gamma}_1=ik\tau+\phi_3(\sigma,\tau), \,\tilde\rho=\frac{\rho}{2}=\tilde\rho(\sigma,\tau),\,\phi=\phi(\sigma,\tau).
\end{eqnarray}
It is instructive to rewrite this solution in Poincare coordinates\footnote{To pass from the embedding coordinates
to the Poincare ones we have used the relations $Y^{\mu}=\frac{y^{\mu}}{z}$,
where $\mu=0,1,2,4$ with the direction 0 playing the role of time and $Y^{-1}+Y^{3}=\frac{1}{z}$,
$Y^{-1}-Y^{3}=\frac{z^2+y^{\mu}y_{\mu}}{z}$. }. It reads:
\begin{eqnarray}\label{EucPoincare}
y^0=a\tanh{k\tau}, \,\,\,z=\frac{a}{\cosh{k\tau}\cosh{\tilde\rho}},\,\,\\ \nonumber
y^1=a\tanh{\tilde\rho}\frac{\cosh{(k\tau+i\phi)}}{\cosh{k\tau}}, \,\,\,y^4=-i a\tanh{\tilde\rho}\frac{\sinh{(k\tau+i\phi)}}{\cosh{k\tau}},\,\,\,{\gamma}_1=ik\tau+\phi_3(\sigma,\tau).
\end{eqnarray}
where $a$ is an overall scale which we have introduced by rescaling
all the Poincare coordinates.

Let us now comment on the solution \eqref{Eucsolution}, \eqref{EucPoincare}.
Firstly, we would like to note that because of the double
Wick rotation of \eqref{analcont}, the target spacetime
is defined by $-(Y^{-1})^2+(Y^{0})^2+(Y^{1})^2+(Y^{2})^2+(Y^{3})^2+
(Y^{4})^2=-1$ and as a consequence it becomes Euclidean $AdS_5$.\\
Secondly, it is easy to see that \eqref{EucPoincare}
describes a string which tunnels from the boundary to the
boundary of the $AdS_5$ space. Indeed, at $\tau=-\infty$
\eqref{EucPoincare} directly gives
\begin{eqnarray}\label{t=-infty}
z=0, \qquad y^0=-a \qquad y^1=a\tanh{\tilde\rho} e^{-i\phi} \qquad y^4=ia\tanh{\tilde\rho} e^{-i\phi}
\end{eqnarray}
which describes a string sitting on the boundary ($z=0$) at $y^0(-\infty)=-a$ and extending along a light-like direction ($(y^1)^2+(y^4)^2=0$).
Similarly, at $\tau=\infty$ we get
\begin{eqnarray}\label{t=infty}
z=0, \qquad y^0=a \qquad y^1=a\tanh{\tilde\rho} e^{i\phi} \qquad y^4=-ia\tanh{\tilde\rho} e^{i\phi}
\end{eqnarray}
which also defines a string sitting on the boundary and extending along another light-like direction.
This tunnelling behaviour of string solution dual to twist $J$ operators has already been studied in \cite{Roiban:2010fe,Georgiou:2010an}.
It should be stressed that the classical solution of \eqref{EucPoincare} does not end at two points on the boundary, as it happens
with solutions which are extended only along coordinates of the five-sphere. However, this fact does not signal a problem regarding
the positions on the boundary where the string vertex operators and thus the SYM dual operators should be inserted.
As shown in \cite{Roiban:2010fe,Georgiou:2010an} these positions are the following points on the boundary
$w_1=(-a,0,0,0)$ and $w_3=(a,0,0,0)$. This implies that the distance between the two operators is given by
\begin{eqnarray}\label{distance}
|w_{13}|=2 a.
\end{eqnarray}

We are now in position to evaluate the three-point structure constant
of two large spin operators dual to the string solution \eqref{EucPoincare} and a
supergravity state, at strong coupling.
For simplicity, we will choose the BPS state to be the BMN vacuum
\begin{eqnarray}\label{CPO}
{\cal O}_I(x)=\frac{1}{\sqrt{l}} \tr {Z^l}(x),
\end{eqnarray}
where $Z=\frac{\Phi_1+i \Phi_2}{\sqrt{2}}$ is one of the complex scalar fields of $N=4$ SYM.
Following the normalisations of \cite{Zarembo:2010rr} the
corresponding spherical harmonic is
\begin{eqnarray}\label{harmonic}
Y_I({\bold n})=\Big( \frac{n_1+i n_2}{\sqrt{2}}\Big)^l=\frac{1}{2^{\frac{l}{2}}} \sin^l{\theta} e^{i l \gamma_1},
\end{eqnarray}
where ${\bold n}$ is a six-dimensional unit vector defining a point on the 5-sphere,
\begin{eqnarray}\label{5sphere}
{\bold n}=n_i=(\sin{\theta}\cos{\gamma_1},\sin{\theta}\sin{\gamma_1},\cos{\theta}\sin{a}\cos{\gamma_2},
\cos{\theta}\sin{a}\sin{\gamma_2},\nonumber\\\cos{\theta}\cos{a}\cos{\gamma_3},
\cos{\theta}\cos{a}\sin{\gamma_3}).
\end{eqnarray}
In \eqref{5sphere} $\gamma_1,\,\gamma_2,\,\gamma_3$ parametrise the three isometries of the sphere.
For all the solutions considered in this note $a=\theta=\pi/2$ which means that $n_2=n_3=n_5=n_6=0$.

The correlators on which we will focus are
\begin{eqnarray}\label{corr1}
<\bar{{\cal O}}_{SJ}(w_3)\,{\cal O}_{SJ}(w_1)\, {\cal O}_I(x)>
\end{eqnarray}
where ${\cal O}_{SJ}$  are the operators dual to the string solutions of \eqref{EucPoincare}.
The OPE coefficient we are after is given by \cite{Zarembo:2010rr}
\begin{eqnarray}\label{CPOstrong}
C_{{\bar{\cal O}}_{SJ} {\cal O}_{SJ}{\cal O}_I} \,|w_{13}|^{\Delta_I}=\frac{2^{{\frac{l}{2}-3}}(l+1) \sqrt{l \lambda}}{\pi N}\int d\tau d\sigma
Y_I({\bold n})z^l(\frac{\partial_a X^{\mu}\partial^a X_{\mu}-\partial_a z\partial^a z}{z^2}-
\partial_a \bold{n}\partial^a \bold{n}).
\end{eqnarray}
In order to proceed we need to evaluate the integrand of \eqref{CPOstrong}. To this end we rewrite it as
\begin{eqnarray}\label{integrand}
H=\frac{\partial_a X^{\mu}\partial^a X_{\mu}-\partial_a z\partial^a z}{z^2}-
\partial_a \bold{n}\partial^a \bold{n}=\frac{\partial_a X^{\mu}\partial^a X_{\mu}+\partial_a z\partial^a z}{z^2}+\partial_a\gamma_1\partial^a\gamma_1\nonumber\\
-2\frac{\partial_a z\partial^a z}{z^2}-\partial_a \bold{n}\partial^a \bold{n}-\partial_a\gamma_1\partial^a\gamma_1.
\end{eqnarray}
On the right hand side of \eqref{integrand}  and in the first line one can easily recognise the Lagrangian density of the bosonic part
of the Polyakov action. It can be most easily evaluated using \eqref{action2Euc} and the second of the Virasoro constraints \eqref{VirasoroEuc}.
 Using \eqref{EucPoincare} one can evaluate the different term appearing in \eqref{integrand} to get
\begin{eqnarray}\label{terms}
\frac{\partial_a X^{\mu}\partial^a X_{\mu}+\partial_a z\partial^a z}{z^2}+\partial_a\gamma_1\partial^a\gamma_1=\frac{1}{2} \big( (\partial_{\sigma}\rho)^2 +\sinh^2{\rho} (\partial_{\sigma}\phi)^2\big)\nonumber\\
\frac{\partial_a z\partial^a z}{z^2}=\frac{k^2 \sinh^2{k \tau}}{\cosh^2{k \tau}}+...\nonumber\\
\partial_a \bold{n}\partial^a \bold{n}=-k^2+...\nonumber\\
\partial_a\gamma_1\partial^a\gamma_1=-k^2+...,
\end{eqnarray}
where the dots in \eqref{terms} denote terms with subleading powers of $k$ which can be ignored. For the same reason the
first line of \eqref{terms} is subleading with respect to the contributions coming from the other terms in \eqref{integrand}.
Overall we obtain
\begin{eqnarray}\label{integrandfinal}
H=-2\frac{k^2 \sinh^2{k \tau}}{\cosh^2{k \tau}}+2 k^2+...=\frac{2k^2}{\cosh^2{k \tau}}+...
\end{eqnarray}
As a result, the structure coefficient becomes \cite{Zarembo:2010rr}
\begin{eqnarray}\label{CPOstrongsemifinal}
C_{{\bar{\cal O}}_{SJ} {\cal O}_{SJ}{\cal O}_I}^{(strong)} \,(2 a)^l=a^l\frac{2^{{\frac{l}{2}-3}}(l+1) \sqrt{l \lambda}}{\pi N}\int_{-\infty}^{\infty} d\tau \int_0^{2 \pi} d\sigma
\frac{1}{2^{l/2}}\frac{e^{i l(i k \tau+\phi_3)}}{\cosh^l{k \tau}\cosh^l{\tilde\rho}}\frac{2k^2}{\cosh^2{k \tau}}.
\end{eqnarray}
Although both ${\tilde\rho}$ and $\phi$ depend on $\tau$ the denominator $1/\cosh^{(l+2)}{k \tau}$ localises the integrand of \eqref{CPOstrongsemifinal}
around $\tau =0$ \cite{Escobedo:2011xw}. The $\tau$ integration can then be easily performed to give
\begin{eqnarray}\label{tauintegration}
\int_{-\infty}^{\infty}d\tau \frac{e^{-l k \tau}}{\cosh^{l+2}{k \tau}}=\frac{2^{l+1}}{k(l+1)}
\end{eqnarray}
Using this result we finally obtain
\begin{eqnarray}\label{CPOstrongfinal}
C_{{\bar{\cal O}}_{SJ} {\cal O}_{SJ}{\cal O}_I}^{(strong)}=\frac{\sqrt{\lambda}k \sqrt{l}}{ N} \int_0^{2 \pi} \frac{d\sigma}{2 \pi}=\frac{J \sqrt{l}}{ N} \int_0^{2 \pi} \frac{d\sigma}{2 \pi}
\frac{e^{i l \phi_3(0,\sigma))}}{\cosh^l{\tilde\rho(0,\sigma)}}.
\end{eqnarray}
This is the final and main result of this Section. In the next Section we will compute the same quantity at weak coupling to find
agreement with \eqref{CPOstrongfinal}.

\section{Weak coupling regime}

In this Section we present the weak coupling computation of the structure constants where two of the operators are non-protected operators
with large spins while the third one is the BMN vacuum. Since the operators we are considering belong to
an $SL(2)$ subsector of the full $PSU(2,2|4)$ superconformal algebra of $N=4$ SYM the orthodox way to proceed would be to find
the Bethe eigenstates of this sector's one-loop dilatation operator, which has been shown to be equivalent to the Hamiltonian of the $XXX_{-\frac{1}{2}}$
spin chain \cite{Beisert:2003jj}, and plug them in the correlator to extract the structure constant.
However, the exact Bethe eigenstates for operators having large values of $S$ and $J$ (see \eqref{twistJop}) are very complicated entangled
quantum states. Fortunately, when the length of the spin chain is large, i.e. in the large $J$ limit a huge simplification occurs.
Namely, the exact eigenstates can be approximated by coherent states \cite{Kruczenski:2003gt,Kruczenski:2004cn}. Although these coherent states are not exact eigenstates of the
1-loop dilatation operator they have the following important property. If one computes with them the average of any classical quantity, such as the energy or the spin, the result one gets agrees with the exact result obtained from the exact
eigenstates up to finite size corrections\cite{Escobedo:2011xw}. Let us mention that the effective low-energy dynamics
of the coherent spin chain states are governed by \eqref{action3} (after removing the total derivative term)\cite{Stefanski:2004cw,Bellucci:2004qr}

The coherent state approach in the $SL(2)$ sector has been studied in \cite{Stefanski:2004cw,Bellucci:2004qr}.
At each site of the spin chain one has a coherent state parametrised by a point on the upper sheet of the two-dimensional
hyperboloid
\begin{eqnarray}\label{parametrise}
{\bold l}=(\cosh{\rho},\sinh{\rho}\sin{\phi},\sinh{\rho}\cos{\phi}),\qquad {\bold l}^2=l_0^2-l_1^2-l_2^2=1,\,\,\,\,l_0>0.
\end{eqnarray}
The $SL((2)$ coherent state is defined by applying the following 'rotation' operator on the lowest weight state $|0\rangle=|\frac{1}{2},\frac{1}{2}\rangle$
of an infinite dimensional irreducible representation of the $SL(2)$ algebra.
\begin{eqnarray}\label{coherent}
|{\bold l}>=e^{\xi J_+ -{\bar \xi}J_-}|0\rangle,\,\,\,\,\,\xi=\frac{\rho}{2}e^{i\phi}.
\end{eqnarray}
In \eqref{coherent} $J_+$ and $J_-$ denote the creation and annihilation operators which together with $J_0$ define the
$SL((2)$ algebra through the commutation relations
\begin{eqnarray}\label{algebra}
[J_-,J_+]=2 J_0, \qquad
[J_0 , J_{\pm}]= \pm J_{\pm},\,\,\,J_{\pm}=J_2\mp i J_1.
\end{eqnarray}
One can show that that the the coherent state can also be expressed in the form \cite{Bellucci:2004qr}
\begin{eqnarray}\label{cohrep}
|{\bold l}>=\frac{1}{\cosh{\frac{\rho}{2}}} \sum_{m=0}^{\infty}e^{i m\phi} \tanh^m{\frac{\rho}{2}}|\frac{1}{2},\frac{1}{2}+m\rangle.
\end{eqnarray}
We list now two important properties of the coherent states. The first one is that they have unit norm.
The second one is that they are not orthogonal.
\begin{eqnarray}\label{properties}
\langle {\bold l}|{\bold l}\rangle=1\nonumber \\
\langle {\bold l_1}|{\bold l_2}\rangle=\frac{1}{\cosh{\frac{\rho_1}{2}}\cosh{\frac{\rho_2}{2}}-\sinh{\frac{\rho_1}{2}}\sinh{\frac{\rho_2}{2}}e^{i(\phi_2-\phi_1)}}
\end{eqnarray}

We are now ready to write down the coherent state representation of the operators
dual to the string solutions considered in the previous Section.
The physical picture is that of a varying spin wave pointing in
the direction $\bold l$ which  is slowly changing from site to site.
The operator $O_1$ will be represented by the following coherent state
\begin{eqnarray}\label{O1}
\langle O_1|=\prod_{i=1}^{J_1}\otimes \langle {\bold l}(\frac{i}{J_1})|.
\end{eqnarray}
Similarly, the second non-protected operator $O_1$ will be given by
\begin{eqnarray}\label{O2}
|O_2\rangle=\prod_{i=1}^{J_2}\otimes | {\bold l}(\frac{i}{J_2})\rangle.
\end{eqnarray}
We should mention that the function ${\bold l}(\sigma)={\bold l}(2 \pi \frac{i}{J})$
appearing in \eqref{O1} and \eqref{O1} is the same. This is the incarnation that $O_1$
is almost the complex conjugate of $O_2$.
Finally the BMN vacuum $\frac{1}{\sqrt{l}}\tr{Z^l}$ can be written as
\begin{eqnarray}\label{BMN}
|O_I\rangle=\frac{1}{\sqrt{l}}\prod_{i=1}^{l}\otimes | {\bold l_0}(\frac{i}{l})\rangle,
\end{eqnarray}
where ${\bold l_0}(\sigma)$ is the constant vector ${\bold l_0}(\sigma)=(1,0,0)$.
This is fully consistent with the fact that the BMN vacuum $\tr{Z^l}$ lives in the centre $\rho=0$ of
the $AdS_5$ space.

A first observation is that the norm of \eqref{O1} and \eqref{O2} is 1 due to the first equation in
\eqref{properties}. A second one is that the R-symmetry imposes the condition
\begin{eqnarray}\label{R}
J_1=J_2+l.
\end{eqnarray}
Notice also that the vectors $\bold l(\frac{i}{J})$ depend also on time
since the spin chain coherent state is dynamical. However, following \cite{Escobedo:2011xw} and taking into account
that in the strong coupling result the main contribution comes from the region around $\tau=0$ we have set $t=0$
to all coherent state vectors appearing in \eqref{O1} and \eqref{O2}. To simplify notation we have suppressed the
$t=0$ argument in $\bold l(\frac{i}{J})$, i.e. $\bold l(\frac{i}{J})=\bold l(\frac{i}{J},t=0)$.
\begin{figure}[!t]
  \centering
  \includegraphics[width=.75\textwidth]{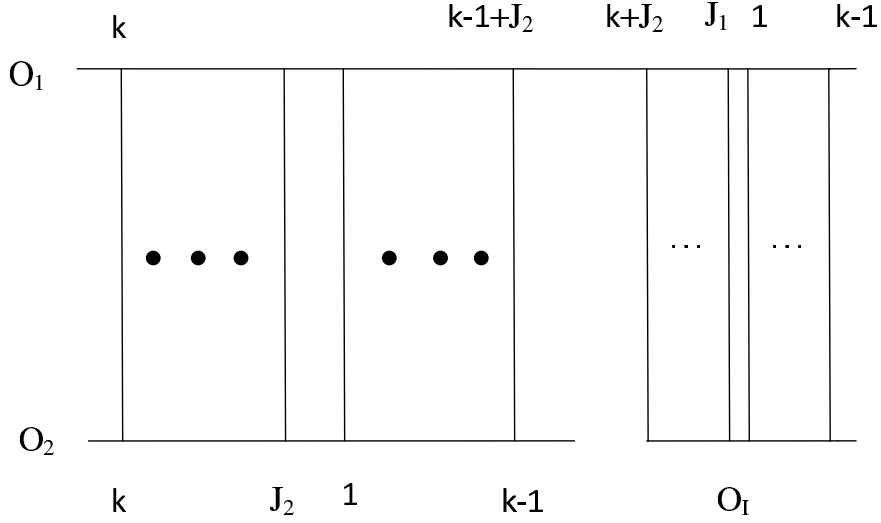}
  \caption{ Tree level contractions contributing to the 3-point structure constant at weak coupling.
  $O_1$ and $O_2$ are large ($J_1\approx J_2>>1$) operators in the $SL(2)$ sector which are almost conjugate
  to each other and are well approximated by coherent states. $O_I$ is the small ($l<<J_1$) BPS operator. $k$ is the starting point
  for the contractions between $O_1$ and $O_2$. R-charge conservation implies that $J_1=J_2+l$.}
  \label{fig:Figure1}
\end{figure}

The three-point structure constant is obtained by Wick contracting the three operators of
\eqref{O1}, \eqref{O2} and \eqref{BMN}. The operator $O_2$ can be Wick contracted with the
'long' operator $O_1$ at sites $k, k+1,..., J_2+k-1$. As a result,  the BPS state $O_I$ should then
be contracted with $O_1$ at sites $J_2+k, J_2+k+1,...,J_1,1,...,k-1$ (see Figure 1).
Finally, one should sum over the insertion starting point $k$.
Let us start by evaluating the Wick contractions between $O_1$ and $O_2$.
These are given by
\begin{eqnarray}\label{O1O2contract}
{\cal I}_k=\prod_{i=k}^{J_2+k-1} \langle {\bold l}(\frac{i}{J_1})|{\bold l}(\frac{i}{J_2})\rangle=\exp{\sum_{i=k}^{J_2+k-1}}
\log{\langle {\bold l}(\frac{i}{J_1})|{\bold l}(\frac{i}{J_2})\rangle}=\nonumber\\
\exp{\int_{\frac{2 \pi k}{J_1}}^{\frac{2 \pi (J_2+k-1)}{J_1}}
(-1)J_1 \frac{d\sigma}{2 \pi}
\log{\big(\cosh{\frac{\rho(\frac{i}{J_1})}{2}}\cosh{\frac{\rho(\frac{i}{J_2})}{2}}-
\sinh{\frac{\rho(\frac{i}{J_1})}{2}}\sinh{\frac{\rho(\frac{i}{J_2})}{2}}e^{i(\phi(\frac{i}{J_2})-\phi(\frac{i}{J_1}))}}\big)}=\nonumber\\
\exp{\int_{\frac{2 \pi k}{J_1}}^{\frac{2 \pi (J_2+k-1)}{J_1}}
(-1)J_1 \frac{d\sigma}{2 \pi}
\log{\big(\cosh{\frac{\rho(\sigma)}{2}}\cosh{\frac{\rho(\sigma\frac{J_1}{J_2})}{2}}-
\sinh{\frac{\rho(\sigma)}{2}}\sinh{\frac{\rho(\sigma\frac{J_1}{J_2})}{2}}e^{i(\phi(\sigma\frac{J_1}{J_2})-\phi(\sigma))}}\big)}\nonumber\\
\end{eqnarray}
In passing from the first to the second line of \eqref{O1O2contract} we have use the second equation in \eqref{properties},
as well as the fact that in the large $J_1$ limit $\sum_i\longrightarrow \int J_1 \frac{d \sigma}{2 \pi}$.
The next step is to evaluate the logarithm appearing in \eqref{O1O2contract}.
To this end we use the expansions
\begin{eqnarray}\label{expansions}
 \cosh{\frac{\rho(\sigma\frac{J_1}{J_2})}{2}}=\cosh{\frac{\rho(\sigma)}{2}}+\frac{1}{2}\sinh{\frac{\rho(\sigma)}{2}}\partial_{\sigma}\rho(\sigma)
 \frac{l\sigma}{J_2}+O(\frac{1}{J_2^2})\nonumber\\
 \sinh{\frac{\rho(\sigma\frac{J_1}{J_2})}{2}}=\sinh{\frac{\rho(\sigma)}{2}}+\frac{1}{2}\cosh{\frac{\rho(\sigma)}{2}}\partial_{\sigma}\rho(\sigma)
 \frac{l\sigma}{J_2}+O(\frac{1}{J_2^2})\nonumber\\
 \phi(\sigma\frac{J_1}{J_2})=\phi(\sigma)+ \partial_{\sigma}\phi(\sigma)
 \frac{l\sigma}{J_2}.
 \end{eqnarray}
 Plugging these expansions in the the logarithm appearing in \eqref{O1O2contract} one gets
\begin{eqnarray}\label{log}
\log{\big(1-i \sinh^2{\frac{\rho(\sigma)}{2}}}\partial_{\sigma}\phi(\sigma)\frac{l\sigma}{J_2}\big)=
-i \sinh^2{\frac{\rho(\sigma)}{2}}\partial_{\sigma}\phi(\sigma)\frac{l\sigma}{J_2}+O(\frac{1}{J_2^2})\nonumber\\
=-i \partial_{\sigma}\phi_3(\sigma)\frac{l\sigma}{J_2}+O(\frac{1}{J_2^2}),
\end{eqnarray}
where in order to get the last line of \eqref{log} we have used the first of the Virasoro constraints \eqref{VirasoroEuc} to
express the derivative of the $AdS_5$ angle $\phi$ in terms of the derivative of the $S^5$ angle $\phi_3$.
Plugging this result in \eqref{O1O2contract} one gets
\begin{eqnarray}\label{O1O2contract2}
{\cal I}_k=\exp{\Big[\int_{\frac{2 \pi k}{J_1}}^{\frac{2 \pi (J_2+k-1)}{J_1}}
J_1 \frac{d\sigma}{2 \pi}  i \partial_{\sigma}\phi_3 \frac{l\sigma}{J_2}}\Big]=\nonumber \\ \exp{\Big[i l \frac{J_1}{J_2 2 \pi}\Big( ( \sigma \phi_3 )|^{2 \pi(\frac{k+J_1-l-1}{J_1})}_{2 \pi(\frac{k}{J_1})}- \int_{\frac{2 \pi k}{J_1}}^{\frac{2 \pi (J_2+k-1)}{J_1}} d\sigma \phi_3
\Big)\Big]}=\nonumber\\
e^{i l \phi_3(2 \pi\frac{k}{J_1})}e^{-i l \Phi}+ O(\frac{1}{J_1}), \qquad  \Phi=\int_{\frac{2 \pi k}{J_1}}^{\frac{2 \pi (J_2+k-1)}{J_1}}
\frac{d\sigma}{2 \pi}\phi_3(\sigma)=\int_{0}^{2 \pi}\frac{d\sigma}{2 \pi}\phi_3(\sigma)+ O(\frac{1}{J_1}).
\end{eqnarray}
Let us now evaluate the result of the contractions between the BPS operator and $O_1$.
The sites of $O_1$ that are contracted are $(J_2+k=J_1-l+k,...,J_1,1,...,k-1)$ (see Figure 1).
These contractions give
\begin{eqnarray}\label{O1BPS}
{\cal J}_k=l \frac{1}{\sqrt{l}}\prod_{i=J_1-l+k}^{k-1}\langle {\bold l}(\frac{i}{J_1})|{\bold l_0}\rangle
\end{eqnarray}
The factor if $1/\sqrt{l}$ appearing in \eqref{O1BPS} is coming from the normalisation of the BPS operator
while the factor of $l$ from the fact that there are $l$ different ways of contracting the BPS operator with
the operator $O_1$. This is so because one can contract any of the $Z$ fields with the first site
$J_2+k=J_1-l+k$ from which the contractions of the BPS state and $O_1$ start.
Furthermore, since the BPS operator is small $l<<J_1$ and the spin wave describing $O_1$ varies slowly
from site to site one can approximate $\langle {\bold l}(\frac{i}{J_1})|{\bold l_0}\rangle$
(up to $1/J_1$ corrections) by its value at the last site $k-1$ or better by its value at site $k$
\begin{eqnarray}\label{approx}
\langle {\bold l}(\frac{i}{J_1})|{\bold l_0}\rangle=\frac{1}{\cosh{\frac{\rho(\frac{k}{J_1})}{2}}}.
\end{eqnarray}
So one gets
\begin{eqnarray}\label{O1BPSfinal}
{\cal J}_k=\sqrt{l}\frac{1}{\cosh^l{\frac{\rho(\frac{k}{J_1})}{2}}}.
\end{eqnarray}
Putting together \eqref{O1O2contract2} and \eqref{O1BPSfinal} we obtain the final result for the three point structure constant at weak coupling.
\begin{eqnarray}\label{CPOweak}
C_{{\bar{\cal O}}_{SJ} {\cal O}_{SJ}{\cal O}_I}^{(weak)}=\frac{1}{N}\sum_{k=1}^{J_1}{\cal I}_k{\cal J}_k=e^{-i l \Phi}\frac{J_1\sqrt{l}}{N}\int_0^{2 \pi}\frac{d\sigma}{2 \pi} \frac{e^{il \phi_3(\sigma)}}{\cosh^l{\frac{\rho(\sigma)}{2}}}.
\end{eqnarray}
By taking into account the $\tilde\rho=\frac{\rho}{2}$ we see that \eqref{CPOweak} is in  agreement with the strong coupling result
\eqref{CPOstrongfinal}
up to an overall phase factor $e^{-i l \Phi}$. This phase factor can also appear in the strong coupling result. When we were writing the
ansatz for the isometry of $S^5$ which is conjugate to the angular momentum $J$ we were having $\gamma_1=i k \tau +\phi_3(\sigma,\tau)$.
But we could very well add any {\it constant} angle to $\gamma_1$ and the equations of motion, as well as the Virasoro constraints would still be
satisfied. Consequently, if we write the solution for $\gamma_1$ as $\gamma_1=i k \tau +\phi_3(\sigma,\tau)-\Phi$ then complete agreement between
the weak and strong coupling structure constants is obtained since the spherical harmonic will give
$e^{i l \gamma_1}=e^{-l k \tau+i l\phi_3(\sigma,\tau)}e^{-il \Phi}$.\\
In any case, as pointed out in \cite{Escobedo:2011xw}, the three-point coefficients are not completely unambiguous even if we canonically
normalise the operators participating
in the correlator. This is so because one can multiply any of these correlators by a constant phase. This phase will not alter the two-point
functions of each of the
operators but generically it will change the structure constant by a phase. \\
Another important comment is that the weak/strong coupling match we have found holds for {\it any} solution in the Frolov-Tseytlin limit.

Some additional comments are in order. The exponential appearing in the weak coupling structure constant
\eqref{CPOweak} should be interpreted through the Virasoro constraint since the $S^5$ angle $\phi_3$ is meaningless from the field theory point of view.
The $\phi_3$ appearing in \eqref{CPOweak} should be understood as the solution of the Virasoro constraint
$2 \partial_{\sigma}\phi_3= (\cosh{\rho}-1)\partial_{\sigma}\phi$
with periodic boundary conditions.\\
A second comment is related to the particular form of the operators used. It is well known that in the case where $J_1=J_2+l$,
i.e. that is in the case where there
are no contraction between operators $O_2$ and $O_I$, the mixing of the single trace operators $O_1$ with double trace operators is important in the
evaluation of the structure constant \cite{Constable:2002hw,Beisert:2002bb}. However, this kind of correlators have been recently
studied in the literature at strong coupling
without taking into account
the mixing with multi-string states. For the same kind of correlators, involving operators in the $SU(3)$ sector,
agreement between the weak and the strong coupling structure constants was found in \cite{Escobedo:2011xw}.
There it was argued that either the same effect ( meaning the effect of mixing with double traces or double string states)
is being forgotten both at weak and strong coupling leaving a remainder quantity that can still be matched or that the effect of the mixing
is suppressed  in the large $J$ limit. The agreement we have found can be viewed from the same perspective.
It is clearly desirable to find a more refined approach where these effects are taken into account.

Another issue is the fate of this agreement when one includes genuine one-loop contributions to the three-point functions.
If both the strong and weak coupling results accept expansions in terms of the quantity $\frac{\lambda}{J^2}$, it would be interesting to see
if some of the corresponding coefficients agree, as it happens with the anomalous dimensions/string energies. This agreement, even if it is there,
should fail at some point due to the well-known by now order of limits problem.
As commented in \cite{Escobedo:2011xw}, this agreement should be, more decently, viewed as a guide to the all-loop result.
Finally, it should be nice to generalise the weak/strong coupling agreement found in the $SU(3)$ sector \cite{Escobedo:2011xw}
and the one found in this note
to the full $PSU(2,2|4)$ algebra.

\section{Appendix}

In this Appendix we are having a closer look at the spacetime structure of the correlators we have considered
in the main text. The careful reader should have noticed that we have treated these correlation function as if they were
scalar. However, this is not true for correlators involving the twist $J$ operators of \eqref{twistJop}.

First of all let us specify the light cone directions along which the derivatives of the operators \eqref{twistJop} are taken.
These directions are determined from the orientation of the string  when it touches the boundary.
By inspecting \eqref{t=-infty} we conclude that the derivatives of the operator ${\cal O}_{SJ}(w_1)$ are taken along the light-like direction
$e_+^{\mu}=(y^0,y^1,y^2,y^4)=\frac{1}{\sqrt{2}}(0,1,0,i)$. In a similar way,
\eqref{t=infty} implies that in its conjugate $\bar{{\cal O}}_{SJ}(w_3)$
the derivatives are taken along $e_-^{\mu}=(y^0,y^1,y^2,y^4)=\frac{1}{\sqrt{2}}(0,1,0,-i)$. By defining the light-like coordinates
$y^+=\frac{y^1+iy^4}{\sqrt{2}}$ and $y^-=\frac{y^1-iy^4}{\sqrt{2}}$ and bearing in mind that the boundary of $AdS_5$ has Euclidean signature
we obtain $\eta_{+-}=\frac{\partial y^{\mu}}{\partial y^+}\frac{\partial y^{\nu}}{\partial y^-}\delta_{\mu\nu}=1$.
Let us first consider the two-point function of ${\cal O}_{SJ}(w_1)$ and its conjugate $\bar{{\cal O}}_{SJ}(w_3)$.
Conformal invariance of the theory dictates the form of the two-point functions of operators which are
symmetric and traceless in the vector indices to be \cite{Fradkin:1998,Chu:2003ji}
\begin{eqnarray}\label{2pointconformalgen}
\langle\bar{{\cal O}}_{\mu_1...\mu_S}(w_3) \, {\cal O}_{\nu_1...\nu_S}(w_1)\rangle=
\frac{ {\rm sym}[J_{\mu_1 \nu_1}(w_{31})J_{\mu_2 \nu_2}(w_{31})...J_{\mu_S \nu_S}(w_{31})]}{w_{13}^{2\Delta}}
\end{eqnarray}
where ''sym'' denotes the symmetrisation and subtraction of traces performed in
each group of indices $\mu_1...\mu_S$, $\nu_1...\nu_S$
and
$J_{\mu\nu}$ is the inversion tensor
\begin{eqnarray}\label{inversion}
J_{\mu\nu}(y)=\delta_{\mu\nu}-2\frac{y^{\mu}y^{\nu}}{y^2}.
\end{eqnarray}
These symmetric and traceless operators belong in irreducible representations of the conformal group in Euclidean space.

The twist $J$ operators that we have consider in this note are exactly of this form.
For the case in hand all $\mu_i=-,\,i=1,...S$ while all $\nu_j=+,\,j=1,...S$.
Furthermore since $w_{31}^{\mu}=(2a,0,0,0)$ we see that $w_{31}$ has no component along either of the
light-cone directions $+$ or $-$, i.e. $w_{31}^{+}=w_{31}^{-}=0$. As a result
\begin{eqnarray}\label{jmn=1}
J_{\mu_i \nu_j}(w_{31})=\eta_{-+}=1
\end{eqnarray}
and our two-point function becomes
\begin{eqnarray}\label{2pointconformal}
\langle \bar{{\cal O}}_{SJ}(w_3) \, {\cal O}_{SJ}(w_1)\rangle=
\frac{(S!)^2}{w_{13}^{2\Delta}}.
\end{eqnarray}

We now turn to the conformal structure of the three-point functions of the form \eqref{corr1}.
To simplify notation, let us consider the case where one of the $SL(2)$ operators has only two
indices, i.e. it is a symmetric tensor of rank two. In this case conformal invariance imposes that the three-point
correlator should have the form \cite{Fradkin:1998}
\begin{eqnarray}\label{corr1conformal}
\langle \bar{{\cal O}}_{\mu_1...\mu_S}^{(\Delta_3)}(w_3)\,{\cal O}_{\mu\nu}^{(\Delta_1)}(w_1)\, {\cal O}_I(w_2)^{(\Delta_2)}\rangle=
\Big[A_1(\lambda)Y^1_{\mu\nu}(w_2 w_3)Y^3_{\mu_1...\mu_S}(w_2 w_1)+\nonumber\\
A_2(\lambda)\Big[ Y^1_{\mu}(w_2 w_3) \frac{1}{w_{31}^2}
\Big(\sum_{k=1}^S J_{\nu \mu_k}(w_{31})
Y^3_{\mu_1...{\hat\mu_k}...\mu_S}(w_2 w_1)-{\rm traces}\Big)+(\mu\leftrightarrow\nu)-{\rm trace \, in \, \mu \, and \, \nu}\Big]\nonumber\\
+A_3(\lambda)\Big[\frac{1}{(w_{31}^2)^2} \sum_{k,r=1}^S J_{\mu \mu_k}(w_{31})J_{\nu \mu_r}(w_{31})
Y^3_{\mu_1...{\hat\mu_k}...{\hat\mu_r}...\mu_S}(w_2 w_1)-{\rm traces}\Big]\Big]D_2(w_1,w_2,w_3),
\end{eqnarray}
where the hat denotes the absence of the corresponding index and $A(\lambda)$, $B(\lambda)$ and $C(\lambda)$ are
coupling dependent constants that can not be determined from just the conformal nature of the theory.
Let us now define the quantities appearing in \eqref{corr1conformal}.
Besides the inversion tensor $J_{\mu\nu}$, a second conformal vector is of importance \cite{Henn:2009bd}.
\begin{eqnarray}\label{Y}
Y^1_{\mu}(w_2 w_3)=\frac{w_{12}^{\mu}}{w_{12}^2}-\frac{w_{13}^{\mu}}{w_{13}^2}.
\end{eqnarray}
This structure is conformally covariant at point $w_1$ and invariant at points $w_2$ and $w_3$.
Finally, the higher order tensors are defined by
\begin{eqnarray}\label{Yn}
Y^1_{\mu_1...\mu_n}(w_2 w_3)=Y^1_{\mu_1}(w_2 w_3)...Y^1_{\mu_n}(w_2 w_3)-{\rm traces},
\end{eqnarray}
while
\begin{eqnarray}\label{D2}
D_2(w_1,w_2,w_3)=\frac{1}{(w_{23}^2)^\frac{\Delta_3+\Delta_2-\Delta_1-S+2}{2}(w_{13}^2)^\frac{\Delta_3-\Delta_2+\Delta_1-S-2}{2}
(w_{12}^2)^\frac{-\Delta_3+\Delta_2+\Delta_1+S-2}{2}}.
\end{eqnarray}

From \eqref{corr1conformal} it is clear how to generalise this expression to the case where the operator at point $w_1$ has $S'<S$
indices.
This three-point correlator should have $S'+1$ different terms which means that to completely determine it one needs
$S'+1$ different constants $A_i, \, i=1,2,...,S'+1$. The first term should be similar to the first term in the right hand side of
\eqref{corr1conformal} (the term multiplying $A_1$ ) except that it should have a tensor structure like
$Y^1_{\nu_1...\nu_{S'}}(w_2 w_3)Y^3_{\mu_1...\mu_S}(w_2 w_1)$.
The second term (the one involving $A_2$) should involve $Y^1_{\nu_1...{\hat \nu_k}... \nu_{S'}}$ and so on.
The only other difference is the definition of
$D_2(w_1,w_2,w_3)$ which becomes
\begin{eqnarray}\label{D2general}
D_2^{SS'}(w_1,w_2,w_3)=\frac{1}{(w_{23}^2)^\frac{\Delta_3+\Delta_2-\Delta_1-S+S'}{2}(w_{13}^2)^\frac{\Delta_3-\Delta_2+\Delta_1-S-S'}{2}
(w_{12}^2)^\frac{-\Delta_3+\Delta_2+\Delta_1+S-S'}{2}}.
\end{eqnarray}

Let us now write down the final form of the three-point function as this becomes for operators with the
characteristics of the string solution of Section 2. Bearing in mind that $w_{13}^+=w_{13}^-=0$ we obtain
\begin{eqnarray}\label{final3point}
\langle \bar{{\cal O}}_{\mu_1=-...\mu_S=-}^{(\Delta_3)}(w_3)\,{\cal O}_{\nu_1=+,...\nu_{S'}=+}^{(\Delta_1)}(w_1)\, {\cal O}_I(w_2)^{(\Delta_2)}\rangle=
\sum_{l=0}^{S'} A_{S'+1-l}(\lambda) \frac{S! S'!}{(S-S'+l)!(S'-l)!}\nonumber\\
\frac{1}{(w_{13}^2)^{S'-l}}\Big( \frac{w_{32}^-}{w_{32}^2}\Big)^{S-S'+l}
\Big( \frac{w_{21}^+}{w_{12}^2} \Big)^l D_2^{SS'}(w_1,w_2,w_3)
\end{eqnarray}
To derive the right hand side of \eqref{final3point} we have used \eqref{jmn=1} to set all J's to one as well as the fact
that $Y^3_{\mu=-}(w_2w_1)=\frac{w_{32}^{\mu=-}}{w_{32}^2}-\frac{w_{31}^{\mu=-}}{w_{31}^2}=\frac{w_{32}^{-}}{w_{32}^2}$ since $w_{13}^-=0$.
Furthermore,  all traces are zero due to the vanishing of traces when both indices are $+$ or $-$.

We conclude that the correlators we are interested in, \eqref{corr1}, are determined by conformal symmetry up to a
huge number of structure constants $A_i,\,i=1,...,S'+1$.
A natural question arises. Which one of these constants have we calculated in the main text and for which have we found
agreement between weak and strong coupling?\\
At weak coupling the answer to this question is obvious. It is $A_{S'+1}$ the one we have computed in Section 3
(the corresponding term in \eqref{corr1conformal} is the term
involving $A_3$). This structure constant $A_{S'+1}$ is the $l=0$ term of \eqref{final3point}.
Notice that  all other terms of \eqref{final3point} include fractions like $\frac{w_{12}^{+}}{w_{12}^2}$ because for all
other terms $l\neq 0$.
The argument is that these fractions can never appear at tree level because
there are no propagators connecting operators at points $w_1$ and $w_2$ (see also Figure 1 \footnote{Please notice that
in Figure 1 $O_1$ denotes the barred operator at point 3 $\bar{{\cal O}}_{\mu_1=-...\mu_S=-}^{(\Delta_3)}(w_3)$, $O_2$
denotes the non-BPS operator at point 2 ${\cal O}_{\nu_1=+,...\nu_{S'}=+}^{(\Delta_1)}(w_1)$, while $O_I$ denotes the BPS vacuum at point 2.}).
Or in other words, $A_{S'+1}=O(\lambda^0)$, whereas $A_{S'}=O(\lambda)$.

Next we argue that it is the same structure constant $A_{S'+1}$ that the method of \cite{Zarembo:2010rr} isolates.
One way to see this is the following. In order to find the structure constant the author of \cite{Zarembo:2010rr}
multiplies the quantity $<{\cal O}_I(w_2)>_W$ by $w_2^{2\Delta_I}$ ($\Delta_I=\Delta_2$) and then
sends $w_2\rightarrow \infty$ keeping $w_1$ and $w_3$ fixed (see equation (2.5) of \cite{Zarembo:2010rr}).
From \eqref{final3point} it is obvious that in this limit it is the $l=0$ term that dominates.
Another way that leads to the same result is the following. One can keep $w_2$ fixed and let both $w_1$ and $w_3$ approach zero. This is so
because the theory is conformal so one can rescale all three points in such a way that brings the large $w_2$ to a finite value.
But then $w_1\approx w_3\approx 0$ which also means that $w_{13}\rightarrow 0$.
By inspecting \eqref{final3point} one can easily see that in the limit $w_{13}\rightarrow 0$ it is the
$l=0$  term (the one which has as coefficient $A_{S'+1}$) that dominates because all other terms behave as the $A_{S'+1}$ term times $|w_{13}|$
to some positive power.
Consequently both at weak and at strong coupling it is the coefficient $A_{S'+1}(\lambda)$ that we have calculated.
It is for this one agreement is found.

Another important comment is that the dominating $l=0$ term of \eqref{final3point} is a scalar quantity in the limit we are considering.
This is so because for the string solution we use, and as a result for the dual operators, it holds that $S=S'$, i.e. the operators at point 1
and 3 have the same $AdS_5$ spin although they have slightly different $S^5$ spins. Then from \eqref{final3point} it is obvious that
$\Big( \frac{w_{32}^-}{w_{32}^2}\Big)^{S-S'+l}
\Big( \frac{w_{21}^+}{w_{12}^2} \Big)^l=1$ for $l=0$. This fact justifies our manipulations in Sections 2 and 3 where we have treated
the three-point function as a scalar quantity.

\vspace{1cm}

\noindent {\large {\bf Acknowledgments}}

\vspace{3mm}

\noindent
We wish to thank Dimitrios Giataganas, Valeria Gili, Chrysostomos Kalousios, Jan Plefka, Rodolfo Russo and George Savvidy  for useful discussions.

\bibliographystyle{nb}
\bibliography{botany}

\end{document}